\documentclass{article}

\PassOptionsToPackage{numbers,compress}{natbib}

\usepackage[preprint]{neurips_2026}
\makeatletter
\renewcommand{\@noticestring}{Preprint. Accepted at NeurIPS 2026.}
\makeatother

\usepackage[utf8]{inputenc}
\usepackage[T1]{fontenc}
\usepackage{url}
\usepackage{booktabs}
\usepackage{tabularx}
\usepackage{amsmath}
\usepackage{amsfonts}
\usepackage{nicefrac}
\usepackage{microtype}
\usepackage{xcolor}
\usepackage{graphicx}
\usepackage{algorithm}
\usepackage{algorithmic}
\usepackage{hyperref}

\newcommand{\framework}{\textsc{Masta}}

\title{MASTA: A Feedback-Scheduled Multi-Agent System for End-to-End Tamarin Modeling and Analysis}

\author{%
  Yongkang Xiao \quad Qiyi Deng \quad Jing Chen\thanks{Corresponding authors.} \quad Min Shi\footnotemark[1] \quad Ju Ma \quad Ruiying Du \\
  School of Cyber Science and Engineering, Wuhan University \\
  {\small\texttt{\{xiaoyongkang,qiyideng,chenjing,itachi,juma\_526,duraying\}@whu.edu.cn}} \\
}

\begin{document}

\maketitle

\begin{abstract}
Tamarin-based formal verification of security protocols requires expert-authored multiset rewriting rules and temporal-logic lemmas.
Translating a natural-language protocol description into a complete formal model with an LLM is hard because syntactic, reachability, and property-level errors compound across the multi-step modeling process,
and a single agent that both authors and self-evaluates cannot reliably localize them.

We present \framework{}, a multi-agent system that addresses these difficulties through three innovations.
First, an \emph{operational}--\emph{evaluative} role decomposition assigns rule modeling to an Architect, property modeling to an Auditor, and pure dispatch to a dedicated Arbiter that holds no authoring duties.
Second, the Arbiter implements a Tamarin-anchored scheduler that dispatches the next agent based on the latest verifier verdict and routes problem-tagged fixes through a modification list,
with a six-class problem taxonomy and an LLM deduplication check preventing routing deadlock.
Third, we construct a benchmark of 10 protocols paired with 139 mutation variants and five fine-grained metrics that jointly assess model correctness ($M_1$--$M_3$) and completeness ($M_4$--$M_5$).

On this benchmark, \framework{} on Qwen~3.5-plus improves over OpenCode on the same backbone by up to $+0.27$ on individual metrics and ranks first on both the correctness and completeness averages,
surpassing Codex (GPT-5.5) and Claude Code (Opus~4.7) despite using a weaker LLM.
Ablations on five protocols confirm that removing the Arbiter collapses property and verification scores to zero.
\end{abstract}

\section{Introduction}
\label{sec:intro}

Security protocols form the backbone of modern networked systems,
and flaws at the design level can have practical impact on deployed systems.
For example, flaws in 5G authentication can enable subscriber tracking~\citep{basin2018formal}, 
while weaknesses in EMV contactless payment protocols can allow transactions to bypass cardholder verification~\citep{basin2021emv}.
Formal verification provides a rigorous method for analyzing protocol security by exhaustively exploring all possible protocol executions under a specified adversary model.
The Tamarin prover~\citep{meier2013tamarin} is a widely used tool for formally verifying security protocols under the Dolev--Yao adversary model~\citep{dolev1983security}.
It has been used to verify protocols such as TLS~1.3~\citep{cremers2017comprehensive}, EMV contactless payments~\citep{basin2021emv}, and 5G~AKA~\citep{basin2018formal}, and has uncovered attacks in these deployed protocols.
However, effective use of Tamarin requires substantial domain expertise:
users must translate an informal protocol description into multiset rewriting rules, 
capture the intended adversary model and cryptographic abstractions, and formalize security requirements as lemmas.
This expertise barrier limits Tamarin's practical reach to a small community of specialists,
reducing the extent to which formal verification is applied to deployed protocols.

Recent advances in large language models~(LLMs) have demonstrated strong code-generation capabilities~\citep{chen2021evaluating,li2022competition,li2023starcoder} and shown emerging potential in formal reasoning tasks, such as theorem proving~\citep{yang2023leandojo,first2023baldur}.
A natural question is \emph{whether LLMs can automate Tamarin-based protocol verification}.
Prompting an LLM to produce a complete Tamarin model faces two challenges.
First, the generated Tamarin model must simultaneously satisfy executability, syntactic legality, and lemma-level semantic correctness,
because Tamarin's domain-specific language imposes structural constraints, including well-sortedness of terms,
correct use of persistent versus linear facts, and proper placement of action facts, that differ fundamentally from general-purpose programming.
Empirical evaluations of LLMs on formal-specification inference tasks~\citep{lecong2025formalbench} report that the conjunction of such constraints degrades model quality.
Second, protocol modeling is an incremental multi-step process in which each added rule may break executable reachability and each added lemma may expose a defect in earlier rules,
so end-to-end generation followed by a single final verification concentrates errors at the last stage and obscures their origin.
A naive remedy that lets a single agent generate each step and self-evaluate it fuses the two activities within one context,
exhibiting the same generator/evaluator coupling found in single-agent self-reflection frameworks~\citep{shinn2023reflexion,madaan2023selfrefine}.
Process-level supervision and a structurally independent evaluator are therefore both required.

We address these challenges with \framework{} (Multi-Agent System for Tamarin-based Analysis).
In \framework{}, an Architect is responsible for modeling the protocol and the adversary as Tamarin multiset rewriting rules,
verifying that each rule preserves executability before proceeding.
An Auditor is responsible for extracting security requirements from the protocol description
and encoding them as formal lemmas.
An Arbiter monitors the Tamarin verification results after each authoring step,
decides which agent should act next,
and routes corrective tasks back to the responsible agent whenever a problem is detected.
This design structurally separates generation from evaluation:
the Arbiter's scheduling decisions are grounded in Tamarin's verdicts rather than the generating agents' own assessment,
eliminating the self-evaluation coupling that undermines single-agent approaches.

Our contributions are summarized as follows.
\begin{itemize}
  \item We introduce an end-to-end LLM-based multi-agent framework for Tamarin-based security protocol verification.
        The framework transforms a natural-language protocol description into a complete Tamarin-compatible formal model under tool-anchored process supervision.
        The generated model comprises multiset rewriting rules that capture both the protocol behavior and the Dolev--Yao adversary capabilities, together with first-order temporal-logic lemmas that formally specify the protocol's security requirements.
  \item We propose a role decomposition in which formal-model authoring is distributed across an Architect and an Auditor, while a dedicated Arbiter handles workflow scheduling exclusively.
        The Arbiter dispatches the next agent based on the latest Tamarin verdict and routes corrective tasks back to the responsible agent.
        This design separates generation from evaluation, eliminating the self-evaluation coupling of single-agent approaches by construction.
  \item We construct a benchmark of 10 security protocols, together with their operational and vulnerability mutations, yielding 149 modeling tasks in total.
        We also design five metrics that jointly assess modeling accuracy
        (whether the generated rules and lemmas agree with expert references)
        and security completeness
        (whether the model correctly reflects protocol modifications and detects introduced vulnerabilities).
\end{itemize}

The full source code, the benchmark, and the experimental logs are released via a public repository~\citep{masta2026repo} to support reproducibility.

The remainder of this paper is organized as follows.
Section~\ref{sec:related} reviews related work on automated protocol formalization,
LLM-based formal methods,
and LLM agents for protocol formalization.
Section~\ref{sec:prelim} introduces the necessary Tamarin background.
Section~\ref{sec:method} presents the \framework{} architecture and the three-agent collaboration mechanism.
Section~\ref{sec:bench} defines the benchmark and the five evaluation metrics.
Section~\ref{sec:exp} reports the main comparison and ablation results,
and Section~\ref{sec:conclusion} concludes.

\section{Related Work}
\label{sec:related}

\paragraph{Automated protocol formalization tools.}
A line of non-LLM work narrows the gap between informal protocol descriptions and the inputs of symbolic protocol verifiers~\cite{meier2013tamarin, cremers2008scyther} by introducing higher-level surface languages.
SAPIC~\citep{kremer2014sapic} compiles applied-pi-calculus processes into Tamarin's multiset rewriting rules,
relieving users of authoring rules directly.
Verifpal~\citep{kobeissi2020verifpal} provides a readable surface syntax that compiles to ProVerif and supports passive/active attacker models.
Alice-and-Bob-style specification languages~\citep{basin2015aliceandbob} let users describe protocols at the message-exchange level,
with automated translation to Tamarin or other backends.
These tools all reduce the symbolic-modeling burden,
but they require users to supply structured semi-formal specifications.

\paragraph{LLM-based formal methods and tool-augmented generation.}
LLM-assisted formal methods include theorem proving over pre-existing formal artifacts~\citep{yang2023leandojo,first2023baldur,wu2024lemur} and evaluation of specification synthesis under controlled mutations~\citep{lecong2025formalbench};
all of these either assume the formal model already exists or evaluate generation in a one-shot setting without verifier feedback.
A complementary line augments LLM agents with external tool signals:
Self-Debug~\citep{chen2023selfdebug} iteratively patches code using compiler and test output,
and SWE-agent~\citep{yang2024sweagent} designs agent-computer interfaces that let an LLM manipulate files and run tests to fix real-world software issues.
Self-correction frameworks such as Reflexion~\citep{shinn2023reflexion} and Self-Refine~\citep{madaan2023selfrefine} use verbal critiques but keep the critic and generator within the same agent.

\paragraph{LLM agents for protocol formalization.}
CryptoFormalEval~\citep{curaba2024cryptoformaleval} benchmarks LLMs on vulnerability detection using Alice-and-Bob notation as input rather than natural language.
AutoSM~\citep{mao2025autosm} generates symbolic models for Tamarin and ProVerif with formal soundness guarantees,
but requires human-in-the-loop interaction and is not fully automated.
\framework{} is fully end-to-end: it takes natural-language protocol descriptions as input and automatically produces complete formal models without human intervention.

\paragraph{Comparison with SAPIC and AutoSM.}
Table~\ref{tab:related} in Appendix~\ref{app:related} contrasts \framework{} with SAPIC and AutoSM.
SAPIC is a translator, so the modeling step we automate remains the user's responsibility.
AutoSM is the closer comparison, and the key difference is autonomy:
its Repairer stage requires user interaction, whereas \framework{} runs end to end using only Tamarin verdicts.
The evaluations also differ in what they can detect.
AutoSM reuses the reference model's lemmas and checks whether they still verify on the generated model,
which corresponds to one of our five metrics ($M_3$, verification consistency).
Because \framework{} generates the lemmas itself,
we also evaluate their structural equivalence to the reference ($M_2$) and the fidelity of the modeled message flow ($M_1$).
In addition, because our benchmark includes 139 mutations,
we can assess whether subtle protocol variations are modeled faithfully ($M_4$, $M_5$),
which a benchmark consisting of one model per protocol cannot capture.

\paragraph{Summary.}
Across all three lines of related work, a common gap remains: no existing system takes a natural-language protocol description and produces a Tamarin-verifiable formal model in a fully automated, process-supervised pipeline.
Automated formalization tools require structured semi-formal input rather than prose.
LLM-based formal methods either assume a pre-existing formal artifact or generate in a one-shot setting without verifier feedback.
Multi-agent and self-correction frameworks do not externalize the evaluative role to an independent agent, leaving generation and assessment within the same context.
\framework{} addresses all three gaps simultaneously by combining natural-language input, a role-decomposed multi-agent architecture, and step-level Tamarin feedback as the grounding signal for cross-agent scheduling and fix-job dispatch.

\section{Preliminaries}
\label{sec:prelim}

\paragraph{Tamarin.}
Tamarin~\citep{meier2013tamarin} specifies a protocol as a set of multiset rewriting rules of the form
\[
[L] \xrightarrow{A} [R],
\]
where $L$ and $R$ are multisets of facts representing the rule's premises and conclusions,
and $A$ is a multiset of action facts recorded on the execution trace.
Linear facts in $L$ are consumed on application; persistent facts are retained.
Security properties are expressed as lemmas: first-order temporal-logic formulas over the sequence of action facts.
For example, a secrecy lemma takes the form
\[
\forall\, x, i.\ \texttt{Secret}(x)\,@\,i \implies \neg \exists j.\ \texttt{K}(x)\,@\,j,
\]
asserting that any value marked $\texttt{Secret}$ at trace position $i$ is never learned by the adversary at any position $j$.
For each lemma, Tamarin returns one of two verdicts: \texttt{proved} or \texttt{falsified}.
For evaluation purposes, we additionally use the label \texttt{timeout} when Tamarin fails to return either verdict within the wall-clock budget,
covering both the explicit cutoff and cases where its proof-search heuristics fail to terminate.

\paragraph{Adversary model.}
Tamarin natively embeds the Dolev--Yao attacker~\citep{dolev1983security},
the standard network adversary who controls the communication channel and can intercept, replay, drop, and compose messages from observed terms.
This baseline is extended on a per-protocol basis to model additional adversary capabilities such as long-term key compromise,
which are expressed as extra Tamarin rules that the Architect adds to the formal model.

\paragraph{Action blocks.}
We introduce \emph{action blocks} as a structured intermediate representation that bridges natural-language protocol descriptions and Tamarin rules.
An action block is attributed to a single participant role and contains an ordered sequence of five operation types: \textsc{Receive}, \textsc{GenerateNonces}, \textsc{Compute}, \textsc{Assertion}, and \textsc{Send}.
The operations within a block are totally ordered in this fixed sequence,
and an action-block sequence $A = (b_1, b_2, \ldots, b_n)$ captures the full message-exchange skeleton of the protocol.
Each block corresponds to one Tamarin multiset rewriting rule.
The formal definition of operation types and the block equality predicate used by $M_1$ are given in Appendix~\ref{app:actionblock}.

\begin{figure}[t]
  \centering
  \includegraphics[width=\linewidth]{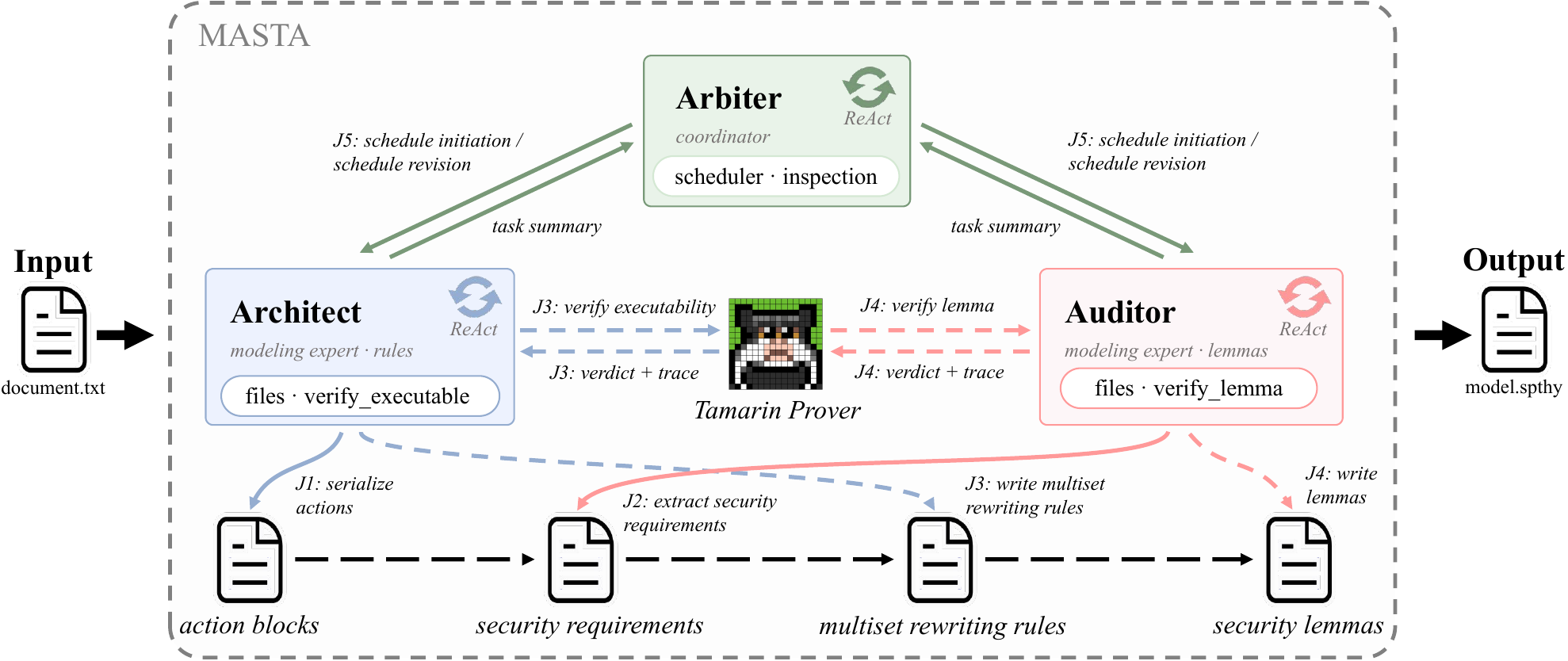}
  \caption{Architectural overview of \framework{}:
    three agents (Architect, Auditor, Arbiter) sharing file I/O and Tamarin invocation;
    four operational jobs (J1--J4) and one continuous scheduling job (J5);
    the Arbiter dispatches the next agent based on the latest Tamarin verdict, including fix-job routing, under a one-hour per-protocol budget.}
  \label{fig:overview}
\end{figure}

\section{Method}
\label{sec:method}

\subsection{Overview}
\label{sec:method:overview}

Figure~\ref{fig:overview} illustrates the overall architecture of \framework{}.
The system processes one target protocol per execution and instantiates three agents: Architect, Auditor, and Arbiter.
Each agent operates on its own system prompt, and all agents share two capabilities: file read/write and Tamarin invocation.
Additionally, only the Arbiter possesses a workflow-scheduling capability.
The system performs five distinct jobs (Table~\ref{tab:jobs}): 
four operational jobs (J1--J4) that generate partial formal models, 
and one continuous scheduling job (J5) that coordinates agents to accomplish the first four jobs.
Input to the agent system is provided as a text file, while the output of each run is a Tamarin formal model consisting of multiset rewriting rules and first-order temporal-logic lemmas. 

\begin{table}[t]
  \centering
  \small
  \caption{The five jobs of \framework{}.
    J1--J4 are operational jobs that produce formal models;
    J5 is a continuous scheduling job whose only output is a dispatch instruction.}
  \label{tab:jobs}
  \begin{tabularx}{\linewidth}{@{}lllX@{}}
    \toprule
    Job & Owner & Input & Output \\
    \midrule
    J1 Action serialization              & Architect & Protocol document        & Structured action blocks \\
    J2 Security-requirement extraction   & Auditor   & Protocol document        & Structured security requirements \\
    J3 Rule modeling                     & Architect & Action blocks            & Tamarin rules + exec.\ verdicts \\
    J4 Property modeling                 & Auditor   & Security requirements    & Tamarin lemmas + lemma verdicts \\
    J5 Workflow scheduling               & Arbiter   & State + verdicts         & Dispatch instruction \\
    \bottomrule
  \end{tabularx}
\end{table}

\subsection{Job Specifications}
\label{sec:method:jobs}

Two intermediate artifacts support the authoring agents.
A \emph{set of action blocks} captures the message-exchange skeleton of the protocol as a sequence of role-attributed receive, fresh-nonce, compute, and send operations, with each action block corresponding to one Tamarin rule template, without committing to Tamarin syntax.
A \emph{structured security requirement} specifies the security assumptions together with the targeted secrecy~\citep{dolev1983security} and authentication~\citep{lowe1997hierarchy} properties between protocol participants.
The four operational jobs are organized so that authoring is interleaved with Tamarin checkpoints, localizing faults to the step that introduced them.

\paragraph{J1 (Architect): action serialization.}
The Architect parses the protocol document into a sequence of action blocks, fixing the message-exchange order before any Tamarin syntax is written.

\paragraph{J2 (Auditor): security-requirement extraction.}
The Auditor parses the same protocol document into a structured security requirements document.
J2 depends only on the protocol document, so it could run in parallel with J1; the default sequence runs it after J1 so that jobs never overlap.

\paragraph{J3 (Architect): rule modeling.}
The Architect translates the action blocks into Tamarin multiset rewriting rules.
After each rule is added, Tamarin is invoked on an exists-trace ``executable'' lemma to confirm that an honest run remains reachable; if the verdict is \texttt{falsified}, the rule is revised before the next rule is added.
This per-rule check ties reachability faults to the rule that introduced them.

\paragraph{J4 (Auditor): property modeling.}
The Auditor writes Tamarin lemmas from the structured security requirements and then issues a single batched Tamarin call once all lemmas are written.
Batching the verification, rather than calling Tamarin per lemma, allows shared restrictions and action-fact conventions to be checked once and amortizes the per-call startup cost.

\paragraph{J5 (Arbiter): workflow scheduling.}
\label{sec:method:scheduling}
The Arbiter is the sole authority over which agent acts next.
On each iteration, it inspects the system state for issues exposed by the latest Tamarin verdict,
selects either the next normal-sequence job or a fix-job for the agent responsible for the affected artifact,
and tags the dispatch with a problem-type label drawn from a fixed six-class taxonomy.
Before issuing the dispatch, the Arbiter runs an LLM deduplication check against the modification list of past dispatches to prevent reissuing the same direction under different wording;
on rejection, a deterministic AdjustDirection procedure cycles through the remaining labels and admissible target artifacts.
A run terminates either with a complete formal model, with a \emph{timeout} when the one-hour budget is exhausted, or with a \emph{stuck state} when AdjustDirection finds no new direction.
The full pseudocode is given as Algorithm~\ref{alg:scheduling} in Appendix~\ref{app:scheduling}.

\subsection{Agent Prompt and Context Design}
\label{sec:method:prompt}

All three agents share a six-layer prompt template, assembled in a fixed order on every turn:

\begin{enumerate}
  \item \textbf{Output Protocol}: the JSON schema for each LLM response.
  \item \textbf{Role Description}: the agent's identity.
  \item \textbf{Capability Description}: the list of callable capabilities.
  \item \textbf{Responsibilities Description}: the agent's per-phase workflow and output formats.
  \item \textbf{Context}: the history of prior \texttt{(action, result)} pairs in the run.
  \item \textbf{Current Task}: the immediate goal set by the Arbiter at each dispatch.
\end{enumerate}

Layers 1--4 are fixed per agent type, while layers 5--6 evolve across turns.
The full template is reproduced in Appendix~\ref{app:prompt}.

\section{Benchmark and Metrics}
\label{sec:bench}

To evaluate \framework{}, we need both a measurement scheme that decomposes model quality into observable, comparable quantities and a benchmark whose inputs and references support that scheme.
This section introduces five metrics ($M_1$--$M_5$) that jointly quantify the correctness and the completeness of a generated Tamarin model (Section~\ref{sec:bench:metrics}),
and describes how the 10-protocol benchmark and its 139 mutation variants were constructed to exercise each of the five metrics (Section~\ref{sec:bench:dataset}).

\subsection{Metrics}
\label{sec:bench:metrics}

\paragraph{Correctness metrics ($M_1$--$M_3$).}
The first three metrics assess whether the generated model is \emph{correct} with respect to the reference.

$M_1$ (process fidelity) captures how faithfully the agent reproduces the protocol's message-exchange structure.
A protocol run is modeled as a sequence of action blocks, each tied to one participant role and containing an ordered list of receive, generate, compute, and send operations.
$M_1$ counts the minimum number of insertions, deletions, and substitutions needed to transform the agent-produced sequence $A$ into the reference sequence $B$, and then normalizes by the length of the longer sequence.
Two action blocks are equal when they share the same role and their operation lists are identical under alpha-equivalence, treating function symbols as constants and allowing only bound variables to be renamed.
$$M_1 = 1 - \frac{\mathrm{EditDistance}(A,\,B)}{\max(|A|,\,|B|)} \in [0,1].$$

$M_2$ (property consistency) captures whether the agent writes security lemmas that are structurally equivalent to the reference lemmas.
Directly checking logical equivalence between two Tamarin lemma formulas is undecidable in general, so we instead provide the agent with a canonical lemma template that fixes quantifier structure, action-fact naming, variable naming, and clause order.
Because the agent is required to follow this template, every generated lemma can be compared to the corresponding reference lemma by direct abstract-syntax-tree matching.
Let the reference contain $N$ lemmas.
Each generated lemma is compared against its reference counterpart and receives a score of 1 if the two are alpha-equivalent as syntax trees and 0 otherwise.
$M_2$ is the mean score over all $N$ reference lemmas.
$$M_2 = \frac{1}{N} \sum_{i=1}^{N} \mathbf{1}[r_i \text{ is alpha-equivalent to its generated counterpart}] \in [0,1].$$

$M_3$ (verification consistency) captures whether running Tamarin on the generated model yields the same verdicts as on the reference model.
A naive binary agreement score is insufficient for \texttt{proved} lemmas, because a vacuously true model in which the protocol is never actually executed would still receive \texttt{proved} from Tamarin.
To penalize such false positives, we use the number of proof steps as a proxy for protocol execution depth.
Let $\hat{v}(l)$ and $v^*(l)$ denote the agent and reference verdicts for lemma $l$, and let $\hat{n}(l)$ and $n^*(l)$ denote the corresponding proof-step counts.
A verdict mismatch receives score 0.
A matching \texttt{proved} verdict receives $\min(1, \hat{n}(l)/n^*(l))$, which is less than 1 when the agent proof uses fewer steps than the reference, indicating a potentially vacuous proof.
All other matching verdicts receive score 1.
$M_3$ is the mean score over all $K$ lemmas.
$$s(l) = \begin{cases} 0 & \hat{v}(l) \neq v^*(l) \\[4pt] \min\!\left(1,\, \dfrac{\hat{n}(l)}{n^*(l)}\right) & \hat{v}(l) = v^*(l) = \texttt{proved} \\[6pt] 1 & \text{otherwise} \end{cases}$$
$$M_3 = \frac{1}{K} \sum_{l=1}^{K} s(l) \in [0,1].$$

\paragraph{Completeness metrics ($M_4$--$M_5$).}
Correctness metrics evaluate the standard-model output but do not test whether the system faithfully captures \emph{variant} protocol behaviors.
Inspired by the mutation-based evaluation of FormalBench~\citep{lecong2025formalbench},
we design two additional metrics to assess model completeness under protocol variation.

$M_4$ (operational-mutation detection) tests whether the agent correctly reflects deliberate but security-preserving changes to the protocol.
Each operational-mutation variant $o$ carries a set of mutation points $\mathcal{P}(o)$, where each point is a subsequence of action blocks that must appear in the generated sequence.
A mutation point $p \in \mathcal{P}(o)$ is considered detected if every action block in $p$ appears in the agent-produced sequence under the $M_1$ equality predicate.
Let $\mathcal{O}$ be the set of all operational-mutation variants of a protocol.
$M_4$ is the fraction of mutation points detected across all variants.
$$M_4 = \frac{\displaystyle\sum_{o \in \mathcal{O}} \sum_{p \in \mathcal{P}(o)} \mathbf{1}[p \text{ is detected}]}{\displaystyle\sum_{o \in \mathcal{O}} |\mathcal{P}(o)|} \in [0,1].$$

$M_5$ (vulnerability-mutation detection) tests whether the agent correctly captures introduced security flaws.
Each vulnerability-mutation variant $u$ is associated with one or more security lemmas whose expected verdict is \texttt{falsified}.
A variant $u$ is considered detected if Tamarin returns \texttt{falsified} on at least one of those lemmas when run on the generated model.
Both a Tamarin \texttt{timeout} and a modeling error that prevents verification count as undetected.
Let $\mathcal{U}$ be the set of all vulnerability-mutation variants of a protocol.
$M_5$ is the fraction of variants correctly detected.
$$M_5 = \frac{1}{|\mathcal{U}|} \sum_{u \in \mathcal{U}} \mathbf{1}[u \text{ is detected}] \in [0,1].$$

\subsection{Dataset}
\label{sec:bench:dataset}

\paragraph{Protocol collection.}
We collected Tamarin formal models from the official Tamarin prover repository~\citep{tamarinproverrepo} and through broader GitHub search,
retaining only models for which Tamarin verification terminates automatically within ten minutes.
For each retained model, we manually authored a natural-language protocol description in the style of an RFC document,
so that the system input is a standalone prose document with no formal content.

\paragraph{Mutation construction.}
For operational mutations, we modify the protocol description to change the message-exchange structure while preserving security, and provide the corresponding reference Tamarin model and action-block sequence.
For vulnerability mutations, we introduce a security flaw into the description and provide the reference model in which at least one security lemma returns \texttt{falsified}.
All variant descriptions and reference files were produced with LLM assistance and verified by a human expert.
The benchmark contains 10 standard protocols together with 101 operational and 38 vulnerability mutation variants, totaling 149 input documents.

\section{Experiments}
\label{sec:exp}

\subsection{Setup}
\label{sec:exp:setup}

We compare \framework{} against three general-purpose coding agents: Codex (GPT-5.5), Claude Code (Opus~4.7), and OpenCode (Qwen~3.5-plus).
\framework{} also uses Qwen~3.5-plus, making the OpenCode--\framework{} pair a controlled comparison where only the agent architecture varies.
All four systems share the same skill set wrapping the four operational jobs (J1--J4) and the same one-hour per-protocol time budget.
Each baseline follows the default sequence J1$\to$J2$\to$J3$\to$J4 and self-evaluates after every job,
revising its own output on failure without dispatching to another agent.
Tamarin version, hardware, and decoding hyperparameters are listed in Appendix~\ref{app:prompt}.
The full source code, the 10-protocol benchmark with all 149 modeling tasks, the per-run experimental logs, and the evaluation scripts that produce the $M_1$--$M_5$ tables below are released in a public repository~\citep{masta2026repo}.

\subsection{Main Comparison}
\label{sec:exp:main}

Table~\ref{tab:main} reports the per-metric $M_1$--$M_5$ scores for every system on the 10-protocol evaluation set.
Because correctness and completeness are not directly additive,
we report Avg(Corr) $=$ mean($M_1$, $M_2$, $M_3$) and Avg(Comp) $=$ mean($M_4$, $M_5$) instead of an unweighted mean over $M_1$--$M_5$.

\begin{table}[t]
  \centering
  \small
  \setlength{\tabcolsep}{4pt}
  \caption{Main comparison on the 10-protocol evaluation set ($n = 10$ protocols).
    Avg(Corr) and Avg(Comp) are the means of $M_1$--$M_3$ and $M_4$--$M_5$, respectively.
    Bold entries indicate the best value per column.}
  \label{tab:main}
  \begin{tabularx}{\linewidth}{@{}XXccccccc@{}}
    \toprule
    System & LLM & $M_1$ & $M_2$ & $M_3$ & Avg(Corr) & $M_4$ & $M_5$ & Avg(Comp) \\
    \midrule
    Codex               & GPT-5.5        & 0.615 & 0.586 & 0.534 & 0.578 & 0.757 & 0.961 & 0.859 \\
    Claude Code         & Opus~4.7       & 0.724 & 0.643 & \textbf{0.594} & 0.654 & \textbf{0.758} & 0.753 & 0.756 \\
    OpenCode            & Qwen~3.5-plus  & 0.762 & 0.430 & 0.526 & 0.573 & 0.668 & 0.866 & 0.767 \\
    \framework{} (Ours) & Qwen~3.5-plus  & \textbf{0.829} & \textbf{0.699} & 0.567 & \textbf{0.698} & 0.745 & \textbf{1.000} & \textbf{0.873} \\
    \bottomrule
  \end{tabularx}
\end{table}

\framework{} ranks first on both sub-averages, with an Avg(Corr) of $0.698$ and an Avg(Comp) of $0.873$,
outperforming every baseline, including those backed by substantially stronger LLMs.
On $M_5$ (vulnerability-mutation detection), it attains a perfect score of $1.000$,
meaning every injected security flaw is correctly exposed as a \texttt{falsified} lemma by Tamarin,
whereas the best baseline (Codex on GPT-5.5) reaches only $0.961$.
The most controlled comparison is OpenCode versus \framework{}: both use Qwen~3.5-plus as the underlying LLM,
so any performance gap is attributable to the agent architecture rather than model capacity.
\framework{} improves over OpenCode by $+0.067$ on $M_1$, $+0.269$ on $M_2$, $+0.041$ on $M_3$, $+0.077$ on $M_4$, and $+0.134$ on $M_5$,
which raises correctness by $0.125$ and completeness by $0.106$,
demonstrating that the three-agent role decomposition and Arbiter-driven scheduling contribute across all evaluation dimensions.
Notably, despite running on Qwen~3.5-plus, \framework{} surpasses Claude Code (Opus~4.7) on $M_1$, $M_2$, and $M_5$,
and surpasses Codex (GPT-5.5) on $M_1$, $M_2$, $M_3$, and both sub-averages,
showing that architectural design can compensate for a weaker underlying model.

\begin{figure}[t]
  \centering
  \includegraphics[width=\linewidth]{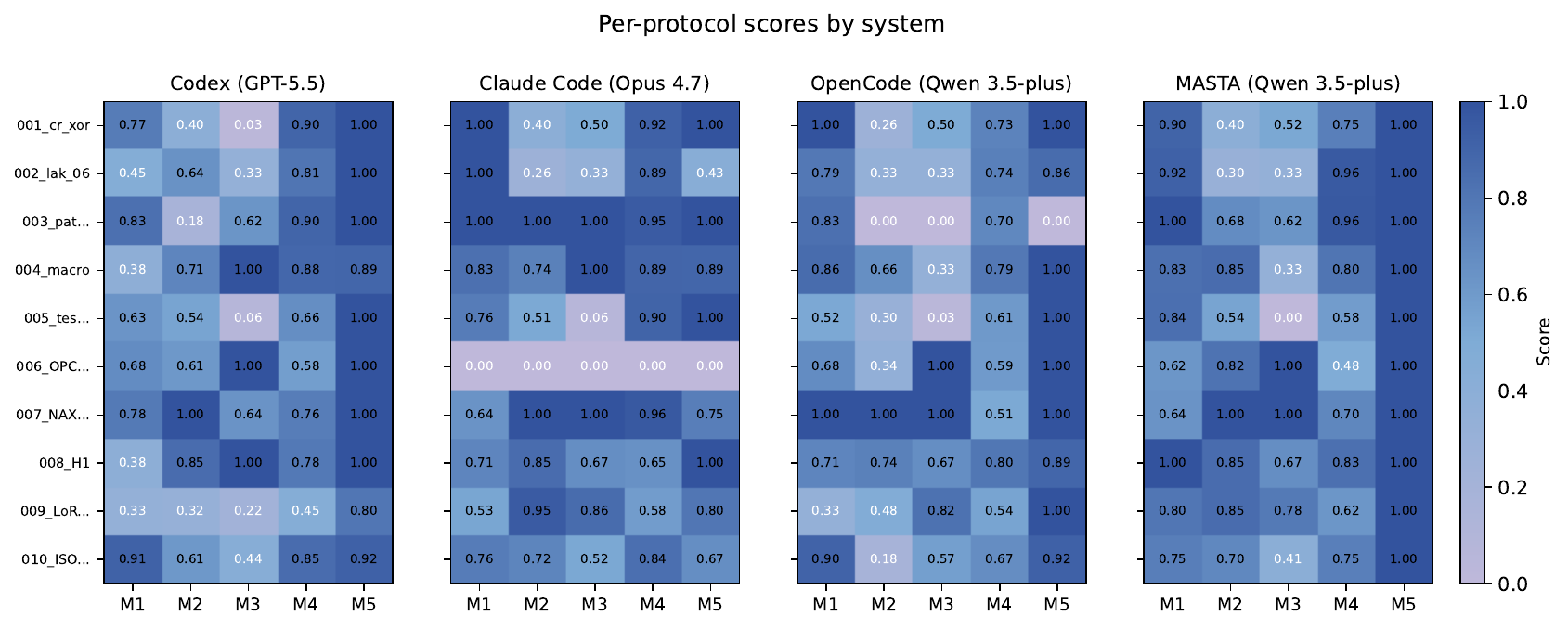}
  \caption{Per-protocol $M_1$--$M_5$ heatmap for the four systems of Table~\ref{tab:main}.
    Rows are the 10 benchmark protocols (sorted by id),
    columns are the five metrics,
    and cells give per-protocol scores.}
  \label{fig:heatmap}
\end{figure}

Figure~\ref{fig:heatmap} breaks these means down by protocol,
making the architectural advantage of \framework{} more concrete.
On $M_5$, \framework{} achieves a perfect score on every single protocol,
confirming that it detects all injected vulnerabilities across the entire benchmark without exception.
On $M_3$, each baseline scores below $0.1$ on two of the 10 protocols,
whereas \framework{} does so only on protocol 005, where it scores $0.00$;
since all four systems score below $0.1$ on that protocol, this entry reflects the difficulty of the protocol rather than a systematic failure of the architecture.

\subsection{Ablation}
\label{sec:exp:abl}

To investigate the contribution of the two core architectural choices in \framework{}, namely the Architect--Auditor operational role split and the dedicated Arbiter, we design two ablation variants.
\emph{w/o Role Specialization} merges the Architect and Auditor into a single operational agent that handles all four jobs, while keeping the Arbiter and its cross-agent scheduling intact.
\emph{w/o Arbiter} retains the Architect--Auditor split but removes the Arbiter entirely, so each agent self-evaluates after every job and revises its own output without any cross-agent dispatch.
Both variants use the same Qwen~3.5-plus backbone and one-hour per-protocol budget as \framework{};
the full configuration of each variant is reproduced in Appendix~\ref{app:abl}.

Table~\ref{tab:ablation} reports the results on five protocols.
\begin{table}[t]
  \centering
  \small
  \setlength{\tabcolsep}{4pt}
  \caption{Ablation results on $n=5$ protocols.
    Differences should be read as descriptive rather than confirmatory given the small scope.}
  \label{tab:ablation}
  \begin{tabularx}{\linewidth}{@{}Xccccccc@{}}
    \toprule
    Variant                  & $M_1$ & $M_2$ & $M_3$ & Avg(Corr) & $M_4$ & $M_5$ & Avg(Comp) \\
    \midrule
    \framework{} (Ours)      & \textbf{0.898} & \textbf{0.553} & \textbf{0.363} & \textbf{0.605} & \textbf{0.811} & \textbf{1.000} & \textbf{0.905} \\
    w/o Role Specialization  & 0.580 & 0.339 & 0.167 & 0.362 & 0.511 & 0.571 & 0.541 \\
    w/o Arbiter              & 0.358 & 0.000 & 0.000 & 0.119 & 0.321 & 0.000 & 0.161 \\
    \bottomrule
  \end{tabularx}
\end{table}
Removing the Arbiter collapses the system: correctness drops by $0.486$ and completeness by $0.744$, with $M_2$, $M_3$, and $M_5$ reduced to zero.
This shows that the Arbiter's cross-agent scheduling and feedback-driven fix-job dispatch are essential for the system to produce a verifiable formal model at all.
Without the Arbiter, each agent's self-evaluation fails to detect and recover from accumulated modeling errors.
Removing the Architect--Auditor split reduces correctness by $0.243$ and completeness by $0.364$.
The gap on $M_4$ and $M_5$ indicates that separating model-side and property-side expertise helps the system handle variant protocols more robustly.

\subsection{Cost and Robustness to Verifier Feedback}
\label{sec:exp:cost}

\paragraph{Verification calls and wall-clock time.}
Across the 149 tasks (10 protocols and 139 mutations), \framework{} averages $10.5$ Tamarin calls and $21$ minutes of wall-clock time per task, for a total of $52.3$ hours.
The feedback loop terminates on the Arbiter's own decision;
the external constraints are a 3,600\,s wall-clock budget per task and a 300\,s budget per Tamarin call.
No task was terminated due to budget exhaustion:
the median wall-clock consumption was 1,058\,s (29\% of the budget),
the longest task took 3,374\,s (93.7\%) and still completed on its own,
and only 5 of the 149 tasks consumed 80\% or more of the budget.

\begin{table}[t]
  \centering
  \small
  \caption{Cost of each system (Tamarin calls, wall-clock time, and API tokens from session logs).}
  \label{tab:cost}
  \begin{tabularx}{\linewidth}{@{}l*{3}{>{\centering\arraybackslash}X}@{}}
    \toprule
    System       & Tamarin calls & Wall-clock time & API tokens \\
    \midrule
    \framework{} & 1,567 & 52.3\,h & 111.1\,M \\
    Codex        & 863   & 8.7\,h  & 81.5\,M \\
    Claude Code  & 966   & 55.4\,h & 184.0\,M \\
    OpenCode     & 3,728 & 39.6\,h & 424.5\,M \\
    \bottomrule
  \end{tabularx}
\end{table}

\paragraph{Cost against baselines.}
Table~\ref{tab:cost} compares the cost of the four systems.
OpenCode, which shares \framework{}'s backbone, makes $2.4$ times as many Tamarin calls (3,728 versus 1,567) and spends $3.8$ times as many tokens, yet scores lower on both sub-averages;
more frequent prover invocation is therefore not what drives the improvement.
Claude Code spends the most wall-clock time and $1.7$ times \framework{}'s tokens, and it also scores lower on both sub-averages.
Codex is the cheapest, using 73\% of \framework{}'s tokens and one sixth of its wall-clock time, but its correctness is $0.120$ below \framework{}'s ($0.578$ versus $0.698$).
\framework{} is the only system that leads on both sub-averages.

\paragraph{Robustness to imperfect verifier feedback.}
Of the 1,567 Tamarin calls, 29 (1.85\%) timed out, and 14 of the 149 tasks (9.4\%) encountered at least one timeout;
93\% of the timeouts were resolved by the responsible agent within its own turn, and all 14 tasks ran to completion.
Appendix~\ref{app:robust} details how the Arbiter handles persistent timeouts, with one representative trace of each Arbiter decision.

\subsection{Second Backbone}
\label{sec:exp:backbone}

To test whether the gains come from the architecture or from synergy with Qwen~3.5-plus,
we evaluate \framework{} with a second, architecturally distinct backbone, Sonnet~4.6.
Because of the time and token cost of a full benchmark run,
we use a reduced subset of three protocols, each with its base model plus 3 operational and 2 to 3 vulnerability mutation variants,
yielding 20 modeling tasks per backbone.
We rerun Qwen~3.5-plus on the identical subset so that the two runs are strictly comparable (Table~\ref{tab:backbone} in Appendix~\ref{app:backbone}).
The two backbones perform comparably:
Sonnet~4.6 is slightly ahead in correctness ($0.747$ versus $0.738$),
while Qwen~3.5-plus is slightly ahead in completeness ($0.978$ versus $0.920$).
We interpret this as evidence that the gains come from the architecture rather than from synergy with a particular backbone.

\section{Conclusion}
\label{sec:conclusion}

We presented \framework{},
an end-to-end multi-agent system that generates executable Tamarin formal models directly from natural-language protocol descriptions.
\framework{} introduces a structural \emph{operational}--\emph{evaluative} role decomposition that distributes formal-model authoring across an Architect and an Auditor and isolates evaluation in a dedicated Arbiter whose only job is workflow scheduling,
together with a dynamic Tamarin-anchored scheduler in which the Arbiter dispatches the next agent based on the latest Tamarin verdict and modeling state and routes problem-type-tagged fixes back to any responsible agent under a deduplicated modification list.
Together, these designs meet the two requirements stated in Section~\ref{sec:intro}: process-level supervision and a structurally independent evaluator.

We also constructed a benchmark of 10 security protocols paired with 101 operational and 38 vulnerability mutations,
totaling 149 modeling tasks,
together with five fine-grained metrics that jointly assess model correctness ($M_1$--$M_3$) and completeness ($M_4$--$M_5$).
On this benchmark, \framework{} achieves the best scores on $M_1$ (process fidelity, $0.829$), $M_2$ (property consistency, $0.699$), and $M_5$ (vulnerability detection, $1.000$) among all evaluated systems,
and ranks first on both the correctness average ($0.698$) and the completeness average ($0.873$) despite using the weaker Qwen~3.5-plus backbone.
The matched-backbone OpenCode--\framework{} pair isolates the contribution of the multi-agent architecture.
Ablations on five protocols confirm that removing the Arbiter collapses $M_2$, $M_3$, and $M_5$ to zero,
while removing the Architect--Auditor split reduces correctness by $0.243$ and completeness by $0.364$,
indicating that the observed gains stem from both the role decomposition and the Arbiter-driven scheduling.

\section*{Limitations}

The limitations of our work are as follows:

\textbf{Benchmark scale and threat model.}
The inclusion criterion (Tamarin auto-termination within ten minutes) caps protocol complexity for tractable $M_1$--$M_5$ evaluation, and the threat model is limited to Dolev--Yao symbolic semantics. 
Larger protocols, where the one-hour budget may be insufficient, and extensions to computational threat models such as CryptoVerif~\citep{blanchet2008cryptoverif} are left for future work.

\textbf{LLM context window.}
The Arbiter's scheduling decision requires the full formal model, the accumulated modification list, and the latest Tamarin verdict to fit within the LLM context window,
so on protocols whose accumulated state exceeds the effective context length, the scheduler may truncate state and miss issues.

\textbf{Tamarin-corpus coverage in pretraining.}
Tamarin-specific syntax generation depends on the LLM's coverage of the Tamarin corpus and degrades on constructs underrepresented in pretraining.
A related concern is data contamination: some benchmark protocols correspond to widely studied designs whose public Tamarin models may overlap with LLM pretraining corpora,
which can inflate baseline scores or compress the gap between systems.
We do not quantify this contamination.

\textbf{Tamarin specialization.}
\framework{} is specialized to Tamarin: prompt templates, role specifications, and the dispatch tag set are all structured around multiset rewriting rules and temporal-logic lemmas,
so migrating to ProVerif or CryptoVerif requires dedicated adaptation while preserving the Architect--Auditor--Arbiter decomposition.
Complementing synthetic mutations with real-world CVE reproduction would also tighten the evaluation against deployed vulnerability distributions.

\begin{ack}
This research was supported in part by the National Natural Science Foundation of China under grants  No. 62625207, 62602479, and the China Postdoctoral Science Foundation - Hubei Joint Support Program under Grant Number 2026T077HB. The  corresponding  authors  are  Jing Chen and Min Shi.
\end{ack}

\bibliographystyle{plainnat}
\bibliography{library}

\appendix

\appendix

\section{Action Block Specification}
\label{app:actionblock}

An \emph{action block} $b = (r,\, \mathit{ops})$ consists of a participant role $r$ and an ordered operation sequence $\mathit{ops} = (o_1, o_2, \ldots)$.
Each operation $o_i$ is drawn from the five types listed in Table~\ref{tab:optype}.
Operations within a block must appear in the fixed total order
\[
\textsc{Receive} \;\prec\; \textsc{GenerateNonces} \;\prec\; \textsc{Compute} \;\prec\; \textsc{Assertion} \;\prec\; \textsc{Send},
\]
and each type may appear zero or more times in that relative order.
If the block corresponds to the final step of a role's participation, the \textsc{Send} entry carries the special marker $\langle\textsc{End}\rangle$ to indicate that no message is transmitted; this marker is used for termination only and does not represent a real network message.

\begin{table}[h]
  \centering
  \small
  \caption{The five operation types inside an action block.}
  \label{tab:optype}
  \begin{tabularx}{\linewidth}{@{}lX@{}}
    \toprule
    Type & Semantics \\
    \midrule
    \textsc{Receive}        & Receive a message from the network; bind its subterms to local variables. \\
    \textsc{GenerateNonces} & Sample fresh random values (nonces, session keys). \\
    \textsc{Compute}        & Apply cryptographic functions (encryption, hash, signature, MAC, DH exponentiation) to produce new terms. \\
    \textsc{Assertion}      & Check a condition on local state (e.g., signature verification, equality test); abort the block if the check fails. \\
    \textsc{Send}           & Transmit a term over the network (or $\langle\textsc{End}\rangle$ if no message is sent). \\
    \bottomrule
  \end{tabularx}
\end{table}

\paragraph{Block equality.}
Two action blocks $b = (r, \mathit{ops})$ and $b' = (r', \mathit{ops}')$ are \emph{equal} iff $r = r'$ and $\mathit{ops}$ and $\mathit{ops}'$ are item-wise alpha-equivalent.
Alpha-equivalence treats all function symbols (encryption schemes, hash functions, etc.) as constants and permits only renaming of bound variables;
in particular, changing a function symbol or reordering operations within the same type counts as a substitution under the $M_1$ edit distance.

\paragraph{Action-block sequence.}
An action-block sequence $A = (b_1, b_2, \ldots, b_n)$ orders the $n$ blocks by their first occurrence in an honest protocol execution trace.
Each block $b_i$ corresponds to exactly one Tamarin multiset rewriting rule in the reference model.

\section{Metric Computation Details}
\label{app:metrics}

This appendix completes the formal specification of the metrics introduced in Section~\ref{sec:bench}.

\paragraph{$M_1$ action-block equality.}
Action-block equality requires identical role and item-wise alpha-equivalent internal operation sequences of the form receive$\rightarrow$nonce$\rightarrow$compute$\rightarrow$send.
Function symbols are treated as constants, and only bound variables may be renamed.
The underlying Levenshtein edit distance charges insertions,
deletions,
and substitutions one unit each.

\paragraph{$M_2$ canonical lemma template.}
Direct logical-equivalence checking on Tamarin lemma formulas is undecidable in general,
so we constrain the agent's output to a canonical template and reduce the equivalence check to AST alpha-equivalence.
The template covers the five property categories in our benchmark---secrecy, aliveness, weak agreement, non-injective agreement, and injective agreement---and fixes for each category:
(i) the quantifier structure (e.g., a fixed prefix \texttt{All A B t \#i.\ ... ==> ...});
(ii) the action-fact vocabulary, drawn from a closed set of facts including \texttt{Create}, \texttt{Running}, \texttt{Commit}, \texttt{Secret}, \texttt{Honest}, and \texttt{LtkReveal} with predefined semantics;
(iii) the variable naming convention; and
(iv) the clause order in the lemma body.
A reference lemma and a generated lemma are then aligned by their position in the lemma list (every protocol's reference fixes the order),
parsed into ASTs,
and compared under alpha-equivalence (function symbols are treated as constants, and only bound variables may be renamed).
A match scores $1.0$, otherwise $0.0$;
generated lemmas violating the template are scored $0.0$ regardless of their content,
and excess generated lemmas beyond the reference list do not contribute to the numerator or the denominator.

\paragraph{$M_3$ proof-step ratio for \texttt{proved} verdicts.}
A purely binary verdict-agreement score gives undue credit to agents whose formal models are vacuously true,
i.e., models under which the protocol is never actually executed and Tamarin trivially returns \texttt{proved} on every safety lemma.
We therefore use the Tamarin proof-step count $n(l)$ as a proxy for proof depth and replace the binary credit on matching \texttt{proved} verdicts with the ratio $\min(1,\, \hat{n}(l) / n^*(l))$,
which is strictly less than $1$ whenever the agent proof terminates in fewer steps than the reference proof.
Verdict mismatches and matching non-\texttt{proved} verdicts are scored as in the main text:
$0$ on mismatch, $1$ on agreement.
The proof-step count is read from Tamarin's standard textual output,
and runs that produce no proof object (e.g., parse errors before invocation) are treated as a verdict of \texttt{error} and contribute $0$ on mismatch with any reference verdict.

\paragraph{Timeout scoring.}
Under the one-hour per-protocol timeout from Section~\ref{sec:method},
$M_1$ and $M_2$ apply the same formulas as non-timeout cases on whatever partial artifacts were produced before the cutoff,
with missing action blocks charged to edit distance and missing lemmas left unmatched.
$M_3$ assigns the agent verdict \texttt{timeout} on every lemma not yet verified;
under the scoring rule above, this collapses to $0$ when the reference verdict is \texttt{proved} or \texttt{falsified},
and to $1$ only in the (rare) case that the reference verdict is also \texttt{timeout}.
A timed-out operational-mutation variant counts in the $M_4$ denominator and contributes zero detected mutation points;
the same convention applies to $M_5$, since a timeout cannot yield a \texttt{falsified} verdict from Tamarin.
Protocol-level aggregation is otherwise unchanged.

\paragraph{No false-positive penalty for $M_4$ / $M_5$.}
$M_4$ false positives are modeling errors already penalized by $M_1$,
$M_5$ false positives are constrained by Tamarin's objective verification verdicts,
and any residual false positive arising from upstream modeling defects is absorbed by $M_1 / M_2 / M_3$,
so $M_4$ and $M_5$ are reported as recall over their respective mutation denominators without an additional false-positive term.

\section{Scheduling Algorithm Details}
\label{app:scheduling}

This appendix reproduces the full pseudocode and the auxiliary deduplication machinery summarized in Section~\ref{sec:method:scheduling}.

\begin{algorithm}[t]
\caption{Arbiter workflow scheduling loop.}
\label{alg:scheduling}
\begin{algorithmic}[1]
\REQUIRE protocol document $D$, time budget $T = 1$ hour
\ENSURE final formal model or timeout
\STATE initialize state $\sigma$ from $D$;
       modification list $L \gets []$
\WHILE{elapsed time $< T$ \AND $\sigma$ is not \emph{Final}}
  \STATE $\mathit{issue} \gets \textsc{DetectIssue}(\sigma)$
         \COMMENT{using latest Tamarin verdicts}
  \IF{$\mathit{issue}$ is \textsc{None}}
    \STATE $\mathit{job} \gets \textsc{NextNormalJob}(\sigma)$
           \COMMENT{J1$\to$J2$\to$J3$\to$J4}
    \STATE $\mathit{target} \gets \textsc{Owner}(\mathit{job})$
  \ELSE
    \STATE $\mathit{job} \gets \textsc{FixJob}(\mathit{issue})$
    \STATE $\mathit{target} \gets \textsc{ResponsibleAgent}(\mathit{issue})$
  \ENDIF
  \STATE $\mathit{dir} \gets (\textsc{ProblemType}(\mathit{job}), \textsc{TargetArtifact}(\mathit{job}))$
  \WHILE{$\textsc{LLMDedup}(\mathit{dir}, L)$ is \TRUE}
    \STATE $\mathit{dir} \gets \textsc{AdjustDirection}(\mathit{dir})$
           \COMMENT{deterministic: cycle problem-type label, then change target artifact}
  \ENDWHILE
  \STATE append $(\mathit{target}, \mathit{dir})$ to $L$
  \STATE $\mathit{output} \gets \mathit{target}.\textsc{Execute}(\mathit{job}, \sigma)$
  \IF{$\mathit{job} \in \{\text{J3},\,\text{J4}\}$}
    \STATE $\mathit{verdict} \gets \mathit{target}.\textsc{RunTamarin}(\mathit{output})$
  \ELSE
    \STATE $\mathit{verdict} \gets \emptyset$
  \ENDIF
  \STATE $\sigma.\textsc{Update}(\mathit{output}, \mathit{verdict})$
\ENDWHILE
\RETURN $\sigma.\textsc{Emit}()$
\end{algorithmic}
\end{algorithm}

\paragraph{Modification-list direction format.}
$\mathit{direction\_text}$ is restricted to a ``problem-type label + target-artifact identifier'' granularity and may not contain a concrete modification operation:
for instance,
``the security requirement disagrees with the protocol description'' is allowed,
whereas ``rename $K$ to $\texttt{ltk}(A,B)$ in the secrecy lemma'' is not.

\paragraph{Six-class problem-type taxonomy.}
The problem-type label is drawn from a fixed six-way taxonomy:
rule unreachable,
property disagrees with protocol,
security-requirement omission,
lemma syntax error,
executable-verification failure,
and other.
Open subtypes are carried by the target-artifact identifier,
so the LLM deduplication check runs over a stable label set.

\paragraph{$\textsc{AdjustDirection}$ procedure and termination.}
Before issuing a new dispatch, the Arbiter calls $\textsc{LLMDedup}(\mathit{dir}, L)$,
which returns true iff the candidate direction is judged equivalent to some entry already in $L$.
On a true verdict, the Arbiter applies $\textsc{AdjustDirection}$,
a deterministic procedure that first cycles through the remaining five problem-type labels in a fixed order,
and only after all six labels have been tried at the current target artifact does it switch the target artifact within the $\mathit{job}$'s admissible artifact set.
The two-level cycle terminates because the direction space is finite,
and the resulting dispatch is guaranteed not to collide with any historical entry in $L$,
which prevents routing deadlock.

\section{Prompt and Agent Configurations}
\label{app:prompt}

This appendix details the six-layer prompt template summarized in Section~\ref{sec:method:prompt}.
Each agent's full prompt is assembled by concatenating the six layers below in the listed order, each rendered under a markdown \texttt{\#\#} section header.
Empty layers are filled with the literal placeholder \texttt{(empty)} so that the LLM never sees a missing section.

\paragraph{Layer 1: {Output Protocol}.}
A ReAct-style behavioral contract that constrains the output of every LLM call to a single JSON object of the form
\begin{verbatim}
{
  "action":  { "capability": "<name>", "args": { ... } },
  "summary": { "did": "...", "last_verification": "...",
               "outstanding": "...", "concern": "..." }
}
\end{verbatim}
Exactly one \texttt{action} is permitted per turn (no arrays).
The four \texttt{summary} fields are mandatory:
\texttt{did} is a one-sentence description of the action and its effect,
\texttt{last\_verification} carries the Tamarin verdict and key error information when the action invoked the verifier (otherwise empty),
\texttt{outstanding} reports what the current phase still needs (e.g., ``the secrecy lemma for the session key is still missing''),
and \texttt{concern} flags any self-detected issue (format ambiguity, repeated failures, missing premises).
The host parses this summary on every turn to decide whether to continue, dispatch a fix-job, or advance to the next phase.

\paragraph{Layer 2: {Role Description}.}
A natural-language statement of the agent's identity and mission, fixed per agent type.
The Architect is positioned as a Tamarin modeling specialist responsible for action blocks and rules;
the Auditor as a cryptographic property specialist responsible for the security requirements and lemmas;
the Arbiter as a workflow coordinator responsible for diagnosing issues and dispatching fixes.

\paragraph{Layer 3: {Capability Description}.}
The capability list is rendered automatically from each agent's capability dictionary,
listing every callable capability together with its signature and calling convention.
The Architect and Auditor share \texttt{files.\{read,write,exists\}} and the Tamarin capabilities \texttt{tamarin.run\_executable} and \texttt{tamarin.verify\_lemma},
plus an \texttt{end} capability that signals task termination.
The Arbiter additionally holds the scheduler capabilities \texttt{scheduler.schedule\_initial}, \texttt{scheduler.schedule\_revision}, and \texttt{scheduler.advance}.
Generating this layer from the live capability dictionary keeps the prompt synchronized with the dispatcher and avoids prompt--code drift.

\paragraph{Layer 4: {Responsibilities Description}.}
A detailed description of the agent's per-phase workflow, expected output artifacts, and failure-handling strategy.
For example, the Architect's J1 phase reads \texttt{document.txt}, writes \texttt{action\_blocks\_draft.yaml}, and terminates;
its J3 phase advances rule by rule, verifying executable reachability after each rule and deciding whether to continue or to revise the most recent rule.
This layer also fixes the canonical output formats (YAML for action blocks, Tamarin for rules and lemmas, markdown for the security requirements) so that the agent's free-form decisions still produce parseable artifacts.

\paragraph{Layer 5: {Context}.}
A serialized history of all prior turns in the current run,
formatted as a sequence of \texttt{(action, result)} pairs separated by horizontal rules.
Each entry records the agent's chosen capability call together with the value returned by the host (file contents, Tamarin verdicts, scheduler responses).
This layer functions as the agent's working memory and ensures that multi-turn reasoning has access to the full execution trace rather than only the most recent step.
A compaction routine bounds context length:
the most recent eight entries are kept verbatim,
while earlier entries are replaced by an LLM-generated summary that preserves verdicts, decisions, and outstanding concerns.

\paragraph{Layer 6: {Current Task}.}
A short, phase-specific instruction that the host (typically the Arbiter) sets at every dispatch.
Examples include ``generate the action blocks from the protocol document and write \texttt{action\_blocks\_draft.yaml}'',
``write a Tamarin lemma for each entry in the security requirements and verify executability'',
and ``inspect the current rule set and choose a target for revision''.
This layer lets the same agent be repurposed across phases without rewriting the upper layers.

\paragraph{Step lifecycle.}
Each invocation of \texttt{agent.step()} runs the following pipeline:
(1) assemble the six-layer prompt;
(2) call the LLM and capture both the thinking-channel output and the JSON answer;
(3) parse the response, validate the \texttt{action} structure and the \texttt{summary} fields, and tolerate stray markdown fences or extra text;
(4) dispatch the action to the corresponding capability handler and append the resulting \texttt{(action, result)} pair to the context;
(5) trigger context compaction if the next prompt would exceed the token budget;
(6) hand the four summary fields to the host, which decides whether to continue, fix, or advance the phase.

\paragraph{Execution environment.}
All experiments use Tamarin prover~1.10.0 with Maude~3.4 and run on a workstation with an AMD Ryzen 9 7950X3D CPU and 192\,GB of memory, executing 16 protocol instances in parallel.
All LLMs are queried with their default API decoding settings.

\section{Reproducibility Materials}
\label{app:repro}

All materials listed below are released via a public repository~\citep{masta2026repo}.
They were frozen before the first experimental run:
\begin{enumerate}
  \item Dataset: the $10$ protocol source documents,
        the standard expert-authored Tamarin reference model of each protocol and the derived reference models for each mutation variant (produced with LLM assistance and verified by a human expert),
        together with the operational-mutation list,
        the vulnerability-mutation list,
        and the position annotations for every protocol.
  \item Benchmark annotations: the reference action-block sequence in JSON and the canonical-format reference lemmas for every protocol and every mutation model.
  \item Interfaces and prompts: the skill signatures, I/O formats, and system prompts of the three agent roles.
  \item Run scripts: the deployment script,
        the ablation-variant switch configuration,
        and the evaluation scripts ($M_1$--$M_5$ computation and protocol-level aggregation).
\end{enumerate}

\section{Ablation Configuration Details}
\label{app:abl}

This appendix reproduces the configuration of the two ablation variants summarized in Section~\ref{sec:exp:abl}.

\paragraph{w/o Role Specialization.}
A single operational expert handles every job previously assigned to the Architect or the Auditor.
The system prompt of this single expert merges the Architect's and the Auditor's role specifications,
and per-job prompt instructions are routed by job type,
so that the model-side jobs (J1 action serialization and J3 rule modeling) reuse the main system's Architect instruction body,
while the property-side jobs (J2 security-requirement extraction and J4 property modeling) reuse the main system's Auditor instruction body.
The Arbiter role,
the workflow scheduling,
and the cross-agent dispatch are all preserved,
except that the dispatch target collapses to the single operational expert;
the one-hour per-protocol timeout and the modification-list deduplication check are unchanged from the main system.
Any drop relative to \framework{} attributable to this variant isolates the contribution of splitting operational work between two specialized experts.

\paragraph{w/o Arbiter.}
The Architect and the Auditor operate without an Arbiter and follow the default job sequence J1$\to$J2$\to$J3$\to$J4 with the same job-to-role assignment as the main system.
Each agent runs Tamarin within its own jobs (executable verification during J3 for the Architect,
batched lemma verification at the end of J4 for the Auditor).
After every job, the current agent self-evaluates using the Arbiter's review prompt rewritten with a ``self-evaluation'' role label,
with the Tamarin verdict (when applicable) and an LLM confirmation of outstanding issues serving as decision rules.
If self-evaluation passes,
control is passed to the next job's designated agent;
if it fails,
the current agent revises its own output without dispatching to any other agent.
The self-evaluation and revision loop shares the main system's one-hour per-protocol timeout.
Upon J4 completion with the Auditor's self-evaluation passing,
the variant emits its final formal model,
which, like the main-system output defined in Section~\ref{sec:method:overview}, consists of Tamarin rules and first-order temporal-logic lemmas;
the sole difference is that the dispatch authority is replaced by the agents' own self-evaluation.
Because the job sequence and the job-to-role assignment are identical to Section~\ref{sec:method:overview},
any performance gap is jointly attributable to the dedicated-evaluator side of the role decomposition and to the dynamic Tamarin-anchored scheduling.

\section{Second-Backbone Results}
\label{app:backbone}

Table~\ref{tab:backbone} reports the per-metric scores of the second-backbone study in Section~\ref{sec:exp:backbone}.

\begin{table}[h]
  \centering
  \small
  \setlength{\tabcolsep}{4pt}
  \caption{\framework{} with two backbones on the same 20-task subset.}
  \label{tab:backbone}
  \begin{tabularx}{\linewidth}{@{}Xccccccc@{}}
    \toprule
    Backbone          & $M_1$ & $M_2$ & $M_3$ & Avg(Corr) & $M_4$ & $M_5$ & Avg(Comp) \\
    \midrule
    Sonnet~4.6        & 0.816 & 0.703 & 0.723 & 0.747 & 0.839 & 1.000 & 0.920 \\
    Qwen~3.5-plus     & 0.967 & 0.643 & 0.604 & 0.738 & 0.956 & 1.000 & 0.978 \\
    \bottomrule
  \end{tabularx}
\end{table}

\section{Comparison with SAPIC and AutoSM}
\label{app:related}

Table~\ref{tab:related} gives the qualitative comparison discussed in Section~\ref{sec:related}.

\begin{table}[h]
  \centering
  \small
  \caption{Qualitative comparison of \framework{} with SAPIC and AutoSM.}
  \label{tab:related}
  \begin{tabularx}{\linewidth}{@{}l>{\raggedright\arraybackslash}X>{\raggedright\arraybackslash}X>{\raggedright\arraybackslash}X@{}}
    \toprule
                        & SAPIC~\citep{kremer2014sapic}                & AutoSM~\citep{mao2025autosm}                                  & \framework{} \\
    \midrule
    Input               & applied-pi process written by the user       & natural-language document                                     & natural-language document \\
    Modeling            & done by the user; SAPIC only translates it   & generated, with user interaction in the Repairer stage        & generated, no human involvement \\
    Repair signal       & not applicable                               & static analysis and user interaction                          & Tamarin verdicts \\
    Security properties & supplied by the user                         & taken from the reference model                                & generated, then verified \\
    Evaluation          & not applicable                               & 18 protocols; a model is correct when the properties verified on the reference still hold on it & 149 tasks; 5 metrics covering process structure, properties, verdicts, and mutation completeness \\
    \bottomrule
  \end{tabularx}
\end{table}

\section{Robustness to Imperfect Verifier Feedback}
\label{app:robust}

This appendix details the timeout analysis summarized in Section~\ref{sec:exp:cost}.
A single Tamarin call has a 300\,s budget, and each task has a 3,600\,s budget.
If a call exceeds its budget, it is retried within the modeling loop;
only exhausting the task budget leaves the task unfinished.
Table~\ref{tab:timeout} reports both measures across all 1,567 calls in our benchmark runs.

\begin{table}[h]
  \centering
  \small
  \caption{Tamarin timeouts across all benchmark runs of \framework{}.}
  \label{tab:timeout}
  \begin{tabularx}{\linewidth}{@{}l>{\raggedright\arraybackslash}X@{}}
    \toprule
    Measure                                  & Value \\
    \midrule
    Total Tamarin calls                      & 1,567 (\texttt{verify\_executable} 744 / \texttt{verify\_lemma} 795 / \texttt{well\_formed} 28) \\
    Timeouts                                 & 29 (1.85\%) \\
    Tasks with at least one timeout          & 14 / 149 (9.4\%) \\
    Tasks terminated by the 3,600\,s budget  & 0 / 149 \\
    \bottomrule
  \end{tabularx}
\end{table}

An isolated timeout is resolved by the agent itself:
it revises \texttt{model.spthy} and reruns the check within the same turn.
This covers most cases.
Persistent timeouts or failures trigger the Arbiter,
which reviews the history and determines whether the current agent is improving in the right direction.
If so, that agent gets additional attempts.
If the fault lies with another role, the Arbiter reassigns the repair to that role and describes what is wrong.
All 14 tasks that encountered timeouts ran to completion.

\paragraph{Arbiter recovery actions.}
The Arbiter recovers through two distinct actions and selects between them based on the fault location:
\texttt{scheduler.advance} grants the current specialist another turn while preserving its context, leaving the assignment unchanged;
\texttt{scheduler.schedule\_revision} cancels the agent while it is still running and re-dispatches the work to the role that owns the damaged artifact,
with a typed problem label and a brief description of the fault.
We provide one trace of each case below, taken directly from our logs.
In both workflows, every lemma receives a definite verdict.

\paragraph{Trace A: the responsible agent is correct, so it is granted more attempts.}
Task \path{007_NAXOS_04_simple_excl3/run_1/action_variants/01}.
The failure occurs in the Architect's rule-modeling stage, the earliest point in the pipeline,
and manifests as two consecutive 300\,s timeouts.
{\footnotesize
\begin{verbatim}
architect  verify_executable -> falsified, timeout (300.12s)
architect  well_formed       -> falsified, timeout (300.25s)
arbiter    concern: "both timed out after 300s, indicating
           structural issues in the theory"
arbiter    -> scheduler.advance      (same Architect continues, context kept)
arbiter    -> scheduler.advance
architect  rewrites model.spthy 3 times (consulting the rule specification)
architect  well_formed       -> verified (253.83s)
architect  verify_executable -> verified (128.76s)
architect  END: "all rules verified"
arbiter    -> schedule_initial(model_properties -> Auditor)
           all 5 security lemmas then obtain clean verdicts
\end{verbatim}
}
After these two consecutive timeouts, the Arbiter did not terminate the task,
nor did it discard the model and reassign the work to another role.
Judging from the verdict history, the Architect was moving in the right direction;
therefore, the Arbiter issued \texttt{scheduler.advance} twice and allowed it to continue.
The timing supports this decision:
\texttt{well\_formed} decreased from a 300\,s timeout to 253.83\,s,
and \texttt{verify\_executable} decreased from a 300\,s timeout to 128.76\,s.
Each rewrite was a genuine improvement, not a blind retry.
The error also remained contained:
the subsequent Auditor stage modeled 5 security lemmas, each of which received a definite verdict.

\paragraph{Trace B: the responsible agent is wrong, so the work is taken away from it.}
Task \path{010_ISO9798_4_4_979844_reuse_once_1_1/run_1/base}.
Here, the Architect had already completed its work correctly, and the error was introduced later by the Auditor.
{\footnotesize
\begin{verbatim}
architect  verify_executable -> verified (0.48s)
architect  END: "all rules verified"
arbiter    -> schedule_initial(model_properties -> Auditor)

auditor    rewrites model.spthy while adding lemmas, but REPLACES the
           Test1()-Test4() action facts instead of appending to them,
           silently breaking the executable lemma that had just passed
auditor    verify_lemma secrecy_key -> falsified (0.44s)
auditor    concern: "violated contract by replacing Test1()-Test4() action
           facts instead of adding to them; executable lemma will fail"

arbiter    -> scheduler.schedule_revision
                target_role     = architect
                target_artifact = model.spthy
                problem_type    = executable-verification failure
           [auditor] cancelled after 5 turns, reason='superseded by new schedule'
architect  start (task_label: revise)
architect  "The Arbiter has flagged an executable-verification failure.
            The brief states that the Auditor replaced Test1()-Test4()..."
architect  16 x verify_executable (0.44-0.60s, no timeouts)
architect  END: "model.spthy now contains Test1()-Test4() alongside
            Running/Commit/Honest/Secret; verify_executable returns verified"

arbiter    -> schedule_initial(model_properties -> Auditor)
auditor    10 x verify_lemma, all clean verdicts (0.51-1.84s)
auditor    END: "all 5 lemmas modeled and verified"
arbiter    END: "workflow complete"
\end{verbatim}
}
The failure appeared in the Auditor's stage, but the defect lay in the rules, which are the responsibility of the Architect.
Following the six-class problem taxonomy (Appendix~\ref{app:scheduling}), the Arbiter mapped this executable-verification failure to the Architect and \texttt{model.spthy}.
It intervened on the first falsified verdict and cancelled the Auditor immediately:
because the Auditor may not edit rules, further lemma-level revisions could not have fixed the defect.
The Architect's first turn restates the brief, naming both the problem type and the specific contract violation,
which confirms that \texttt{schedule\_revision} transmits actionable guidance rather than merely performing a hand-off.

\end{document}